\documentclass{new_tlp}

\usepackage{amssymb}

\usepackage{microtype}
\usepackage{hhline}
\usepackage{stmaryrd}
\usepackage{lmodern}
\usepackage{multirow}
\usepackage{xspace}
\usepackage{textcomp}
\usepackage[pdftex,dvipsnames]{xcolor}
\definecolor{eminence}{RGB}{108,48,130}
\definecolor{weborange}{RGB}{200,130,0}
\usepackage{listings}
\usepackage{amsmath}
\newcommand{\prettylstciao}[0]{
\lstset{language=Prolog,
  frameround=tttt,
  frame=lt,
  rulecolor=\color{blue},
  numbers=left,numberstyle=\tiny,stepnumber=1,numbersep=8pt,
  tabsize=4,
  showstringspaces=false,
  breaklines=true,breakatwhitespace=true,
  showlines=true,
  showspaces=false,showtabs=false,
  upquote=true,
  commentstyle=\color{gray},
  keywordstyle=\color{eminence},
  basicstyle=\footnotesize\ttfamily,
  keywordstyle=\color{weborange},
  emphstyle={\color{blue}},
  emph={pred,prop,trust,check,checked,true,rsize,cardinality,
  not_fails,module,exp,cost,steps_ub,steps_lb,size_ub,size_lb,
  covered,mut_exclusive,cost,use_module,int,calls,success,cost_center,
  is_det,num,nat,var,list,ground,length,terminates,steps_o,resource,
  entry,impl_defined,regtype},
  otherkeywords={>,<,>=,=<,.,;,-,!,=,~,*,\&,+,:-,[,],|,->,:,:=},
  morekeywords= {>,<,>=,=<,.,;,-,!,=,~,*,\&,+,:-,[,],|,->,:,:=},
  escapeinside=\`\`,
}}
\usepackage[colorlinks=true,citecolor=blue]{hyperref}

\newcommand{\secbeg}{\vspace*{-3mm}}
\newcommand{\secend}{}

\title[An Approach to Static Performance Guarantees for Run-time
       Checking]{%
      \textbf{\textit{\underline{\Large Technical report CLIP-1/2018.0}}} \\
      \vspace{1em}
      An Approach to Static Performance Guarantees\\ for
       Programs with Run-time Checks\
      \thanks{This research has received funding from EU
      FP7 agreement no 318337 \emph{ENTRA}, Spanish MINECO
      TIN2015-67522-C3-1-R \emph{TRACES} project, and the Madrid
      M141047003 \emph{N-GREENS} program.}%
      \footnote{An extended abstract of this work is published
      as~\cite{rtchecks-cost-2018-iclp-tc}}}

\author[M. Klemen  et al.]{
       MAXIMILIANO KLEMEN$^{1,2}$
   ~~  NATALIIA STULOVA$^{1,2}$ \and
       \hspace{-15mm}PEDRO L\'OPEZ-GARC\'IA$^{1,3}$ 
   ~~  JOS\'{E} F. MORALES$^{1}$ 
   ~~  MANUEL V. HERMENEGILDO$^{1,2}$
\ \\
\ \\
   $^1$IMDEA Software Institute \\
\email{\{maximiliano.klemen,nataliia.stulova,pedro.lopez,josef.morales,manuel.hermenegildo\}@imdea.org} \vspace*{1mm} \\
   $^2$ETSI Inform\'{a}ticos, Universidad Polit\'{e}cnica de Madrid (UPM) \\
   $^3$Spanish Council for Scientific Research (CSIC) \\
}

\newcommand{\ciao}{Ciao\xspace}
\newcommand{\ciaopp}{CiaoPP\xspace}

\newtheorem{example}{Example}

\newcommand{\kbd}[1]{\mbox{\tt #1}}

\newcommand{\chkarg}[2]
  {\textcolor{blue}{\textsf{\textit{#1}}(\textcolor{black}{#2})}}
\newcommand{\chk}[1]{\textcolor{blue}{\textsf{\textit{#1}}}}

\newcommand{\clause}[2]
  {\ensuremath{#1\textsf{~:-~} #2}}

\newcommand{\fmtFloatTwo}[1]{$#1$}

\newcommand{\fmtRel}[3]%
{\FPeval\Trel{round((#1+#2)*100/#3:0)}\numprint[\%]\Trel}

\definecolor{check}{RGB}{0,0,150}
\definecolor{true}{RGB}{0,150,0}
\definecolor{false}{RGB}{150,0,0}
\definecolor{checked}{RGB}{0,100,0}

\definecolor{lightgrey}{rgb}{0.95,0.95,0.95}

\newcommand{\admissibleoverhead}{admissible run-time checking overhead\xspace}
\newcommand{\admissibleoverheadcap}{Admissible Run-time Checking Overhead\xspace}
\newcommand{\admisoverheadacronym}{\textsf{AOvhd}\xspace}

\newcommand{\overhead}{run-time checking overhead\xspace}
\newcommand{\overheadacronym}{\textsf{Ovhd}\xspace}

\newcommand{\benchmarks}{\textbf{Bench.}}
\newcommand{\colexp}{\textbf{Bound Inferred}}
\newcommand{\colrtc}{\textbf{RTC}}
\newcommand{\coltimes}{$\mathbf{T_A(ms)}$}
\newcommand{\coldev}{\textbf{\%D}}
\newcommand{\colover}{\textbf{\overheadacronym}}
\newcommand{\colaover}{\textbf{Verif.}}

\begin{document}

\maketitle

\begin{abstract}
Instrumenting programs for performing run-time checking of properties,
such as regular shapes, is a common and useful technique that helps
programmers detect incorrect program behaviors.
This is specially true in dynamic languages such as Prolog. 
However, such run-time checks inevitably introduce run-time overhead (in
execution time, memory, energy, etc.).
Several approaches have been proposed for reducing such overhead, such
as eliminating the checks that can statically be proved to always
succeed, and/or optimizing the way in which the (remaining) checks are
performed.
However, there are cases in which it is not possible to remove all
checks statically (e.g., open libraries which must check their
interfaces, complex properties, unknown code, etc.)
and in which, even after optimizations, these remaining checks still
may introduce an unacceptable level of overhead.  It is thus important for
programmers to be able to determine the additional cost due to the
run-time checks and compare it to some notion of admissible cost.
The common practice used for estimating
run-time checking overhead is profiling, which is not
exhaustive by nature. Instead, we propose a method that uses static
analysis to estimate such overhead, with the advantage that the
estimations are functions parameterized by input data sizes. 
Unlike profiling, this approach can provide guarantees for all
possible execution 
traces, and allows assessing how the overhead grows as the size of the
input grows. Our method also extends an existing assertion verification
framework to express ``admissible'' overheads, and statically and
automatically checks whether the instrumented program conforms with
such specifications. 
Finally, we present an experimental evaluation of our approach
that suggests that our method is feasible and promising.
\end{abstract}

\begin{keywords}
  Run-time Checks, Assertions,
  Abstract Interpretation,
  Resource Usage Analysis.
\vspace*{-3mm}
\end{keywords}


\secbeg
\section{Introduction and Motivation}
\label{sec:intro}
\secend

Dynamic programming languages are a popular programming tool for many
applications, due to their flexibility.
They are often the first choice for web programming, prototyping, and
scripting.
The lack of inherent mechanisms for ensuring program data manipulation
correctness (e.g., via full static typing or other forms of full
static built-in verification) has sparked the evolution of flexible
solutions, including
assertion-based approaches~\cite{aadebug97-informal-short,%
  prog-glob-an-shorter,%
  Lai00-short,%
  mod-ctchecks-lpar06-short}
in (constraint) logic languages,
\emph{soft-}~\cite{cartwright91:soft_typing-short,%
  TypedSchemeF08-short}
and \emph{gradual-}~\cite{Siek06gradualtyping}
typing in functional languages,
and contract-based
approaches~\cite{DBLP:journals/fac/LeavensLM07,%
  lamport99:types_spec_lang,nguyen-icfp14-short}
in imperative languages.
A trait that many of these approaches share is that some parts of the
specifications may be the subject of \emph{run-time checking} (e.g.,
those cannot be discharged statically in the case of systems that
support this functionality).
However, such run-time checking comes at the price of overhead during
program execution, that can affect execution time, memory use, energy
consumption, etc., often in a
significant way~\cite{DBLP:conf/popl/RastogiSFBV15-short,%
  DBLP:conf/popl/TakikawaFGNVF16-short}.
If these overheads become too high the whole program execution becomes
impractical and programmers may opt for sacrificing the checks to keep
the required level of performance.

Dealing with excessive run-time overhead is a challenging
problem. Proposed approaches in order to address this problem include
discharging as many checks as possible via static
analysis~\cite{assert-lang-disciplbook-short,%
  assrt-theoret-framework-lopstr99-shorter,%
  clousot-2010-shorter,%
  Hanus17PRELOPSTR,%
  optchk-journal-scp}, optimizing the dynamic checks
themselves~\cite{rv2014-short,cached-rtchecks-iclp2015-shorter,%
  Ren:StatDynChk:PLDI16-short}, or limiting run-time checking
points~\cite{testchecks-iclp09-short}.
Nevertheless, there are cases in which a number of checks cannot be
optimized away and must remain in place, because of software
architecture choices (e.g., the case of the external interfaces of
reusable libraries or servers), the need to ensure a high level of
safety
(e.g., in safety-critical systems), etc.

At the same time, run-time checks may not always be the culprit of low
program performance.
A technique that can help in this context is \emph{profiling}, often
used to detect performance ``hot spots'' and guide program optimization.
Prior work on using profiling in the context of optimizing the
performance of programs with run-time
checks~\cite{druby-oopsla09-short,profiling-padl11-short,%
racket-profiling-cc15-short}
clearly demonstrates the benefits of this approach.
Still, profiling infers information that is valid only for some
particular input data values (and their execution traces).
The profiling results thus obtained may not be valid for other input
data values.  Since the technique is not exhaustive by nature,
detecting the worst cases can take a long time, and is impossible in
general.

Our proposal is to use
\emph{static cost
  analysis}~\cite{granularity-short,%
  caslog-short,%
  low-bounds-ilps97-short,%
  rhlv-02-short-alt,%
  AlbertAGPZ12,%
  plai-resources-iclp14-short,%
  gen-staticprofiling-iclp16-short%
} instead of (or as a complement to)
dynamic profiling.
Such analysis is aimed at inferring statically \emph{safe upper and
  lower bounds on execution costs}, i.e., bounds that are guaranteed
and will never be violated in actual executions.  Since such costs are
data-dependent, these bounds take the form of functions that depend on
certain characteristics (generally, data sizes) of the inputs to the
program. They show how the program costs change as the size of the
input grows.
We propose a static cost analysis-based approach that delivers
guarantees on the costs introduced by the run-time checks in a
program (\overhead).
Our method provides the programmer with feedback at compile-time
regarding the impact that run-time checking will have on the program
costs.
Furthermore, we propose an assertion-based mechanism that allows
programmers to specify bounds on the \admissibleoverhead
introduced in programs.  The approach then compares the inferred
\overhead against the admissible one
and provides guarantees on whether such
specifications are met or not. Such guarantees can be given as
constraints (e.g., intervals) on the size of the input data.
We provide the formalization of the method in the context of the \ciao
assertion language and the \ciaopp verification framework, and present
also results from its implementation and experimental evaluation.


\secbeg
\section{Assertions and Run-time Checking}
\label{sec:run-time-checking}
\secend

\paragraph{Assertion Language}
Assertions are linguistic constructions that allow expressing properties
of programs.
We recall the \kbd{pred} assertions of the Ciao
assertion language~\cite{%
  prog-glob-an-shorter,%
  assert-lang-disciplbook-short,%
  hermenegildo11:ciao-design-tplp-shorter},
following the presentation of~\cite{optchk-journal-scp}.
Such \kbd{pred} assertions allow defining 
the set of all admissible preconditions for
a given predicate, and for each %
such pre-condition a corresponding post-condition.
These pre- and post-conditions are formulas containing literals defined
by predicates specially labeled as \emph{properties}, to which we refer
to as \emph{prop} literals.
A set of
assertions for a predicate (identified by a normalized atom $Head$) is:
\begin{small}
  \[
  \begin{array}{l}
    \kbd{:- } Status \kbd{ pred } Head \kbd{ : } Pre_1 \kbd{ => } Post_1 \kbd{.}
  \\
    \ldots
  \\
    \kbd{:- } Status \kbd{ pred } Head \kbd{ : } Pre_n \kbd{ => } Post_n \kbd{.}
  \end{array}
  \]
\end{small}
The $Pre_i$ and $Post_i$ fields are logic formulas (e.g., conjunctions)
of \emph{prop} literals that refer to the variables of $Head$.
Informally, such a set of assertions states that in any execution state
immediately before the call to $Head$
at least one of the $Pre_i$ conditions
should hold, and that, given the $(Pre_i,Post_i)$ pair(s) where
$Pre_i$ holds, then, if the predicate succeeds, the corresponding
$Post_i$ should hold upon its success.

\begin{example}[Program with Assertions]
\label{ex:running-code}
\noindent%
\emph{%
Consider the following implementation of a predicate for reversing a
list and its assertions:}\\ [-6mm]
\begin{center}
\begin{minipage}[c]{0.98\textwidth}
\prettylstciao
\begin{lstlisting}
:- check pred rev(X,Y) : (list(X),var(Y)) => (list(X),list(Y)).  % A1
rev([], []).     rev([X|Xs], Y):- rev(Xs, Ys), app(Ys,X,Y).

:- check pred app(Y,X,Z) : (list(Y), term(X), var(Z))    % \ A2
                        => (list(Y), term(X), list(Z)).  % /
app([],X,[X]).   app([E|Y],X,[E|T]):- app(Y,X,T).
\end{lstlisting}
\end{minipage}
\end{center}
\vspace*{-3mm}
\emph{%
Assertion \kbd{A1} states that if \kbd{rev/2} is called with
a list \kbd{X} and a free variable \kbd{Y},
on its success \kbd{Y} 
will also be a list.
Assertion \kbd{A2} says if \kbd{app/3} is called with a list \kbd{Y},
a term \kbd{X},
and a free variable \kbd{Z}, on success 
\kbd{Z} will be a list.}
\emph{The algorithmic complexity of \kbd{rev/2} 
  is $O(N^2)$ in the size (list length in this case) $N$ of its input
  argument \kbd{X}.
While this implementation is obviously not optimal, we use it as a
representative of the frequent case of nested loops with linear costs.
}
\vspace*{-2mm}
\end{example}

Every assertion also has a $Status$ field which indicates whether the
assertion refers to intended or actual properties.
Programmer-provided assertions by default have status \kbd{check}, and
only assertions with this status generate run-time checks.
Static analysis can prove or disprove properties in assertions for a
given class of input queries, statically verifying assertions (if all
the prop literals proved to be true, in which case their status is
changed to \kbd{checked}) or flagging errors (if any prop literal is
proved to be false, and then the status is changed to \kbd{false}).
Assertions can also be simplified by eliminating the prop literals
proved to be true, so that only the remaining ones need to be checked.
Other information inferred by static analysis is communicated by means
of \kbd{true} assertions (e.g., see Ex.~\ref{ex:running-assrts-cost}).

\begin{example}[Assertions Processed by Static Analysis]
\label{ex:running-assrts}

\begin{center}
\begin{minipage}[c]{0.98\textwidth}
\prettylstciao
\begin{lstlisting}
:- check   calls rev(X,Y) : (list(X),var(Y)).
:- checked pred  rev(X,Y) : (list(X),var(Y)) => (list(X),list(Y)).

:- checked pred  app(Y,X,Z) : (list(Y), term(X), var(Z))
                           => (list(Y), term(X), list(Z)).
\end{lstlisting}
\end{minipage}
\end{center}

\noindent%
\emph{%
Result of static assertion checking for 
Example~\ref{ex:running-code}. We assume that the code is in a module,
exporting only \kbd{rev/2}, and that it is analyzed
in isolation, i.e., we have no information on the callers to \kbd{rev/2}.
The interface assertion (\kbd{calls}) for the \kbd{rev/2} predicate
remains active and generates run-time checks (calls into the module
are sanitized). This 
contrasts 
with 
the situation in Example~\ref{ex:running-code}, where all assertions
generate run-time checks.}

\end{example}

\paragraph{Run-time Check Instrumentation}

We recall the definitional source
transformation of~\cite{cached-rtchecks-iclp2015-shorter},
that introduces \emph{wrapper} predicates that check calls and success
assertions, and also groups all assertions for the same predicate
together to produce optimized checks.
Given a program, for every predicate $p$ the transformation
replaces all clauses $p(\bar{x}) \leftarrow body$ by $p'(\bar{x})
\leftarrow body$, where $p'$ is a new predicate symbol, and inserts
the wrapper clauses given by $\textsf{wrap}(p(\bar{x}),p')$:\\ [-5mm]
\begin{minipage}{\textwidth}
\[
  \textsf{wrap}(p(\bar{x}),p') =
    \left\{
    \begin{array}{l}
    \clause{p(\bar{x})}{p_C(\bar{x},\bar{r}),p'(\bar{x}),p_S(\bar{x},\bar{r})}.\\
    \clause{p_C(\bar{x},\bar{r})}{\mathit{ChecksC}}.\\
    \clause{p_S(\bar{x},\bar{r})}{\mathit{ChecksS}}.
    \end{array}
    \right\}
\]
\end{minipage}

\vspace*{2mm}
\noindent%
Here $\mathit{ChecksC}$ and $\mathit{ChecksS}$ are the
optimized compilation of pre- and postconditions
$\bigvee_{i=1}^{n}Pre_i$ and $\bigwedge_{i=1}^{n}(Pre_i \rightarrow
Post_i)$ respectively;
and the additional \emph{status} variables $\bar{r}$ are used to
communicate the results of each $Pre_i$ evaluation to the
corresponding $(Pre_i \rightarrow Post_i)$ check,
thus avoiding double evaluation of preconditions.

The compilation of checks for assertions emits a series of
calls to a \linebreak \kbd{reify\_check(P,Res)} predicate, which accepts
as the first argument a property \kbd{P} and unifies 
\kbd{Res} with
\kbd{1} or \kbd{0}, depending on whether the property check succeeds or
not.
The results of those reified checks are then combined and evaluated as
boolean algebra expressions using bitwise operations and the
Prolog \kbd{is/2} predicate.
That is, the logical operators $(A \vee B)$, $(A \wedge B)$, and $(A
\rightarrow B)$ used in encoding assertions are replaced by
their bitwise logic counterparts \kbd{R is A \char`\\/ B},
\kbd{R is A /\char`\\
  ~B}, \kbd{R is (A \# 1) \char`\\/ B}, respectively.

\begin{example}[Run-time Checks (a)]
\label{ex:running-rtchecks-orig}
\emph{%
The program transformation that introduces the run-time checking harness
for the program fragment from Example~\ref{ex:running-code}
(assuming none of the assertions has been statically discharged by 
analysis) is essentially as follows:}
\vspace*{-2mm}

\begin{center}
\begin{minipage}[c]{0.95\textwidth}
\prettylstciao
\begin{lstlisting}
rev(A,B) :-                       app(A,B,C) :-
    revC(A,B,C),                      appC(A,B,C,D),
    rev'(A,B),                        app'(A,B,C),
    revS(A,B,C).                      appS(A,B,C,D).

revC(A,B,E) :-                    appC(A,B,C,G) :-
    reify_check(list(A),C),           reify_check(list(A),D),
    reify_check(var(B), D),           reify_check(term(B),E),
                                      reify_check(var(C), F),
    E is C/\D,                        G is D/\(E/\F),
    warn_if_false(E,'calls').         warn_if_false(G,'calls').

revS(A,B,E) :-                    appS(A,B,C,G) :-
    reify_check(list(A),C)            reify_check(list(A),D),
    reify_check(list(B),D),           reify_check(term(B),E),
                                      reify_check(list(C),F),
    F is C/\D, G is (E#1)\/F,         H is D/\(E/\F), K is (G#1)\/H,
    warn_if_false(G,'success').       warn_if_false(K,'success').

rev'([],[]).                      app'([],X,[X]).
rev'([X|Xs],Y) :-                 app'([E|Y],X,[E|T]) :-
    rev(Xs,Ys),app(Ys,X,Y).           app(Y,X,T).
\end{lstlisting}
\end{minipage}
\end{center}
\emph{%
The \kbd{warn\_if\_false/2} predicates raise run-time
errors
terminating program execution if their first argument is \kbd{0}, and 
succeed (with constant cost) otherwise.
We will refer to this case as to the \emph{worst} performance case in
programs with run-time checking.}
\end{example}

\begin{example}[Run-time Checks (b)]
\label{ex:running-rtchecks-orig-interface}

\begin{center}
\begin{minipage}[c]{0.42\textwidth}
\prettylstciao
\begin{lstlisting}
rev(A,B) :-
    revC(A,B),
    rev'(A,B).

revC(A,B) :-
    reify_check(list(A),C),
    reify_check(var(A), D),
    E is C/\D,
    warn_if_false(E,'calls').

rev'(A,B) :- rev_i(A,B).

rev_i([],[]).
rev_i([X|Xs],Y) :-
    rev_i(Xs,Ys),app(Ys,X,Y).
\end{lstlisting}
\end{minipage}
\hfill
\begin{minipage}[c]{0.55\textwidth}
\vspace*{-4mm}  
\emph{%
This example represents the run-time checking generated for the
scenario of Example~\ref{ex:running-assrts}, i.e., after applying
static analysis to simplify the assertions.
Run-time checks are generated only for the interface calls of the
\kbd{rev/2} predicate.
Note that \kbd{rev'/2} here is a point separating calls to \kbd{rev/2}
coming \emph{outside} the module from the \emph{internal} calls (now
made through \kbd{rev\_i/2}).
Note also that \texttt{app/3} is called directly (i.e., with no
run-time checks). 
Clearly in this case there are fewer
checks in the code and thus smaller overhead.
We will refer to this case, where only interface checks remain, as the
\emph{base} performance case.}
\end{minipage}
\end{center}
\end{example}


\secbeg
\section{Static Cost Analysis}
\label{sec:static-analysis}
\secend

Static cost analysis automatically infers information about the
resources that will be used by program executions, without actually
running the program with concrete data.
Unlike profiling, static
analysis can provide guarantees (upper and lower bounds) on the
resource usage of all possible execution traces, given as functions on
input data sizes.  In this paper we use the \ciaopp general cost
analysis
framework~\cite{granularity-short,low-bounds-ilps97-short,%
resource-iclp07-short,plai-resources-iclp14-short},
which is parametric with respect to \emph{resources},
\emph{programming languages}, and other aspects related to cost.
It can be easily customized/instantiated by the user to infer a wide
range of resources~\cite{resource-iclp07-short}, including resolution steps,
execution time, energy consumption,
number of calls to a particular predicate,
bits sent/received by an application over a socket, etc.

In order to perform such customization/instantiation, the \ciao
assertion language is
used~\cite{resource-iclp07-short,assert-lang-disciplbook-short,%
  hermenegildo11:ciao-design-tplp-shorter}.
For cost analysis it allows
defining different resources and how basic components of a
program (and library predicates) affect their use.
Such assertions constitute the \emph{cost model}.
This model is taken (trusted) by the static analysis engine, that
propagates it during an abstract interpretation of the
program~\cite{plai-resources-iclp14-short} through code segments,
conditionals, loops, recursions, etc., mimicking the actual execution
of the program with symbolic ``abstract'' data instead of concrete
data.
The engine is fully based on
\emph{abstract interpretation}, and defines the resource analysis
itself as an \emph{abstract domain} that is integrated into the PLAI
abstract interpretation framework~\cite{ai-jlp-short} of \ciaopp.

The engine infers cost functions (polynomial, exponential, logarithmic,
etc.) for higher-level entities, such as procedures in the program.
Such functions provide upper and lower
bounds on resource usage that depend on input data sizes and possibly
other (hardware) parameters that affect the particular resource.
Typical size metrics include actual values of numbers, lengths of
lists, term sizes (number of constant and function symbols),
etc.~\cite{resource-iclp07-short,plai-resources-iclp14-short}.
The analysis of recursive procedures sets up recurrence equations (cost
relations), which are solved (possibly safely approximated), obtaining
upper- and lower-bound (closed form) cost functions.
The setting up
and solving of recurrence relations for inferring closed-form
functions representing bounds on the sizes of output arguments and the
resource usage of the predicates in the program are integrated into
the PLAI framework as an abstract operation. \\

\begin{example}[Static Cost Analysis Result]
\label{ex:running-assrts-cost}
\emph{The following assertion is part of the output of the resource
  usage analysis performed by \ciaopp for the \kbd{rev/2} predicate
  from
Example~\ref{ex:running-code}:}
\begin{center}
\begin{minipage}[c]{0.95\textwidth}
\prettylstciao
\begin{lstlisting}
 :- true pred rev(X,Y) : (list(X),var(Y),length(X,L))
                      => (list(X),list(Y), length(X,L), length(Y,L))
                       + cost(exact(0.5*(L)**2+1.5*L+1), [steps]).
\end{lstlisting}
\end{minipage}
\end{center}
\noindent\emph{
It includes, in addition to the precondition (\kbd{:$Pre$}) and
postcondition (\kbd{=>$Post$}) fields,
a field for \emph{computational} properties 
(\kbd{+$Comp$}), in this case cost.}
\emph{The assertion uses the \kbd{cost/2} property for expressing
  the \kbd{exact} cost (first argument of the property) in terms of
  resolution \kbd{steps} (second argument) of any call to
  \kbd{rev(X,Y)} with the \kbd{X} bound to a list and \kbd{Y} a free
  variable. Such cost is given by the function $0.5 \ L^{2} + 1.5
  \ L + 1$, which depends on $L$, i.e., the length of the (input)
  argument \kbd{X}, and is the argument of the \kbd{exact/1} qualifier.
  It means that such function is both a lower and an upper bound on
  the cost of the specified call.
This aspect of the assertion language (including the \kbd{cost/2}
property)
and our proposed extensions are discussed
in Section~\ref{sec:estimate-rtc-overhead}.
}
\end{example}


\secbeg
\section{Analyzing and Verifying the Run-time Checking Overhead}
\label{sec:estimate-rtc-overhead}
\secend

Our approach to analysis and verification of run-time checking
overhead consists of three basic components: using static cost
analysis to infer upper and lower bounds on the cost of the program
with and without the run-time checks;
providing the programmer with a means for specifying the amount of
overhead that is admissible; and comparing the inferred bounds to
these specifications. The following three sections outline these
components.

\secbeg
\subsection{Computing the \overhead (\overheadacronym)}
\label{sec:computing-overhead}
\secend

The first step of our approach is to infer upper and lower bounds on
the cost of the program with and without the run-time checks, using
cost analysis.
The inference of the bounds for the program without run-time checks
was illustrated in Example~\ref{ex:running-assrts-cost}.  The
following two examples illustrate the inference of bounds for the
program with the run-time checks. They cover the two scenarios discussed
previously, i.e., with and without the use of static analysis to
remove run-time checks.

\begin{example}[Cost of Program with Run-time Checks (a)]
\label{ex:running-rtchecks-cost}
\emph{The following is the result of cost analysis for the run-time
  checking harness of Example~\ref{ex:running-rtchecks-orig} for the
  \kbd{rev/2} predicate, together with a (stylized) version of the
  code analyzed, for reference.
  Note the jump in the execution cost of \kbd{rev/2} from quadratic
  to cubic in the size $L$ of the input (the list length of
  \kbd{A}), which is most likely not admissible: }
\begin{center}
\begin{minipage}[c]{0.98\textwidth}
\prettylstciao
\begin{lstlisting}
:- true pred rev(A,B) : (list(A),var(B)) => (list(A),list(B),length(A,L))
                      + cost(exact(0.5*L**3+7*L**2+14.5*L+8), [steps]).

rev(A,B)    :- revC(A,B,C),rev'(A,B),revS(A,B,C).
revC(A,B,C) :- `\chkarg{list}{A}`,`\chkarg{var}{B}`,`\chk{bit\_ops}`.   revS(A,B,C) :- `\chkarg{list}{A}`,`\chkarg{list}{B}`,`\chk{bit\_ops}`.

rev'([],[]).   rev'([X|Xs],Y) :- rev(Xs,Ys),app(Ys,X,Y).
\end{lstlisting}
\end{minipage}
\end{center}
\end{example}

\begin{example}[Cost of Program with Run-time Checks (b)]
\label{ex:running-rtchecks-cost-interface}

\begin{center}
\begin{minipage}[c]{0.6\textwidth}
\prettylstciao
\begin{lstlisting}
:- true pred rev(A,B)
    : (list(A),var(B)) => (length(A,L))
    + cost(exact(0.5*L**2+2.5*L+7), [steps]).

rev(A,B)  :- revC(A,B),rev'(A,B).
revC(A,B) :- `\chkarg{list}{A}`,`\chkarg{var}{B}`,`\chk{bit\_ops}`.

rev'(A,B) :- rev_i(A,B).

rev_i([],[]).
rev_i([X|Xs],Y) :- rev_i(Xs,Ys),app(Ys,X,Y).
\end{lstlisting}
\end{minipage}
\begin{minipage}[c]{0.38\textwidth}
\vspace*{1mm}
  \noindent\emph{%
    This example shows the result of cost analysis for the base
    instrumentation case of
    Example~\ref{ex:running-rtchecks-orig-interface}: although there
    are still some run-time checks present for the interface, the
    overall cost of the \kbd{rev/2} predicate remains quadratic, which
    is probably \emph{admissible}.  }
\end{minipage}
\end{center}
\end{example}

\secbeg
\subsection{Expressing the \admissibleoverheadcap (\admisoverheadacronym)}
\label{sec:express-admissible-overhead}
\secend

We add now to our approach the possibility of expressing the
\admissibleoverhead (\admisoverheadacronym).  This is done by means of
an extension to the \ciao assertion language. As mentioned before,
this language already allows expressing a wide range of properties,
and this includes the properties related to resource usage. For
example in order to tell the system to check whether an upper bound on
the cost, in terms of number of resolution steps, of a call
\kbd{p(A, B)}
with \kbd{A} instantiated to a natural number and \kbd{B} a free
variable, is a function in $O($\kbd{A}$)$,
we can write the following assertion:

\begin{center}
\begin{minipage}[c]{0.98\textwidth}
\prettylstciao
\begin{lstlisting}
:- check pred p(A, B): (nat(A), var(B)) + cost(o_ub(A), [steps, std]).
\end{lstlisting}
\end{minipage}
\end{center}

\noindent 
The first argument of the \kbd{cost/2} property is a cost function,
which in turn appears as the argument of a qualifier expressing the
kind of approximation. In this case, the qualifier \kbd{o\_ub/1}
represents the complexity order of an upper bound function (i.e., the
``big $O$'').  Other qualifiers include \kbd{ub/1} (an upper-bound cost
function, not just a complexity order), \kbd{lb/1} (a lower-bound cost
function), and \kbd{band/2} (a cost band given by both a lower and
upper bound). The second argument of the \kbd{cost/2} property is a list of
qualifiers (identifiers). The first identifier expresses the resource,
i.e., the cost metric used. The value \kbd{steps} represents the
\emph{number of resolution steps}.
The second argument expresses the particular kind of 
cost used. The value \kbd{std} represents the \emph{standard cost}
(the value by default if it is omitted), the value \kbd{acc} the
accumulated cost~\cite{gen-staticprofiling-iclp16-short}, etc.

The language also allows writing assertions that are universally
quantified over the predicate domain (i.e., that are applicable to all
calls to all predicates in a program), which is particularly useful in
our application.  An issue that appears in this context is that
different predicates can have different numbers and types of
arguments. To solve this problem we introduce a way to express
complexity orders without requiring the specification of details about
the arguments on which cost functions depend nor the size metric used,
by means of identifiers without arguments, such as \kbd{constant}, \kbd{linear},
\kbd{quadratic}, \kbd{exponential}, \kbd{logarithmic}, etc.  For
example, in order to extend the previous assertion to all possible
predicate calls in a program (independently of the number and type of
arguments), we can write:

\begin{center}
\begin{minipage}[c]{0.98\textwidth}
\prettylstciao
\begin{lstlisting}
:- check pred * + cost(so_ub(linear), [steps]).
\end{lstlisting}
\end{minipage}
\end{center}

\noindent 
Alternatively, we also introduce complexity order expressions that do
not specify the relation with the arguments of the predicates, i.e.,
we allow the use of free variables in the expressions:

\begin{center}
\begin{minipage}[c]{0.98\textwidth}
\prettylstciao
\begin{lstlisting}
:- check pred * + cost(so_ub(N**4), [steps]).
:- check pred * + cost(so_ub(2**N), [steps]).
\end{lstlisting}
\end{minipage}
\end{center}

In the context of the previous extensions, our objective is expressing
and specifying limits on how the complexity/cost changes when run-time
checks are performed, i.e., expressing and specifying limits on the
run-time checking overhead. To this end we propose different ways to
quantify this overhead. Let ${\cal C}_p(\bar n)$ represent the
standard cost function of predicate $p$ without any run-time checks
and ${\cal C}_{p\_rtc}(\bar n)$ the cost function for the
transformed/instrumented version of $p$ that performs run-time checks,
$p\_rtc$.  A good indicator of the relative overhead is the ratio:

\begin{minipage}{\textwidth}
$$\frac{{\cal C}_{p\_rtc}(\bar n)}{{\cal C}_p(\bar n)}$$
\end{minipage}
\vspace*{0.5mm}

\noindent
We introduce the qualifier \kbd{rtc\_ratio} to express this type of
ratios. For example, the assertion:

\begin{center}
\begin{minipage}[c]{0.98\textwidth}
\prettylstciao
\begin{lstlisting}
:- check pred p(A, B) : (nat(A), var(B)) 
                      + cost(so_ub(linear), [steps, rtc_ratio]).
\end{lstlisting}
\end{minipage}
\end{center}
\vspace*{-1mm}
\noindent
expresses that \kbd{p/2} should be called with the first argument
bound to a natural number and the second one a variable, and the
relative overhead introduced by run-time checking in the calls to \kbd{p/2}
(the ratio between the cost of with and without run-time checks)
should be at most a linear function. Similarly, using the universal
quantification over predicates, the following assertion:

\vspace*{-1mm}
\begin{center}
\begin{minipage}[c]{0.98\textwidth}
\prettylstciao
\begin{lstlisting}
:- check pred * + cost(so_ub(linear), [steps, rtc_ratio]).
\end{lstlisting}
\end{minipage}
\end{center}

\noindent 
expresses that, for all predicates in the program, the ratio between
the cost of with and without run-time checks should be at most a
linear function.  

\secbeg
\subsection{Verifying the \admissibleoverheadcap (\admisoverheadacronym)}
\label{sec:check-admissible-overhead}
\secend

We now turn to the third component of our approach: \emph{verifying
  the \admissibleoverhead (\admisoverheadacronym)}. 
To this end, we leverage the general framework for
resource usage analysis and
verification of~\cite{resource-verif-iclp2010-short,resource-verif-2012-short},
and adapt it for
our purposes, using the assertions introduced in 
Section~\ref{sec:express-admissible-overhead}. The \emph{verification}
process 
compares the (approximated) intended semantics of a program (i.e., the
specification) with approximated semantics inferred by static
analysis. These operations include the comparison of arithmetic
functions (e.g., polynomial, exponential, or logarithmic functions)
that may come from the specifications or from the analysis results.
The possible outcomes of this process are the following:

\begin{enumerate}
 
\item The status of the original (specification) assertion (i.e., 
  \texttt{check}) is changed to \texttt{checked}
  (resp. \texttt{false}), meaning that the assertion is correct
  (resp. incorrect) for all input data meeting the precondition of the
  assertion,

\item the assertion is ``split'' into two or three assertions with
  different status (\texttt{checked}, \texttt{false}, or
  \texttt{check}) whose preconditions include a conjunct expressing
  that the size of the input data belongs to the interval(s) for which
  the assertion is correct (status \texttt{checked}), incorrect
  (status \texttt{false}), or the tool is not able to determine
  whether the assertion is correct or incorrect (status
  \texttt{check}), or

\item in the worst case, the assertion remains with status
  \texttt{check}, meaning that the tool is not able to prove nor to 
  disprove (any part of) it.

\end{enumerate}

In our case, the specifications express a band for the
\admisoverheadacronym, defined by a lower- and an upper-bound cost
function (or complexity orders). If the lower (resp. upper) bound is
omitted, then the lower (resp. upper) limit of the band is assumed to
be zero (resp. $\infty$).

This implies that we need to perform some adaptations with respect to
the verification of resource usage specifications for predicates
described
in~\cite{resource-verif-iclp2010-short,resource-verif-2012-short}.
Assume for example that the user wants the system to check the
following assertion:

\begin{center}
\begin{minipage}[c]{0.98\textwidth}
\prettylstciao
\begin{lstlisting}
:- check pred p(A, B): (nat(A), var(B)) 
                     + cost(ub(2*A), [steps, rtc_ratio])
\end{lstlisting}
\end{minipage}
\end{center}

\noindent 
which expresses that the ratio defined in
Section~\ref{sec:express-admissible-overhead} (with $\bar n =$ A)
$\frac{{\cal C}_{p\_rtc}(\bar n)}{{\cal C}_p(\bar n)}$ must be in the
band $[0, 2*A]$ for a given predicate \texttt{p}. The approach
in~\cite{resource-verif-iclp2010-short,resource-verif-2012-short} uses
static analysis to infer both lower and upper bounds on ${\cal
  C}_p(\bar n)$, denoted ${\cal C}_{p}^{l}(\bar n)$ and ${\cal
  C}_{p}^{u}(\bar n)$ respectively. In addition, in our application,
the static analysis needs to infer, both lower and upper bounds on
${\cal C}_{p\_rtc}(\bar n)$, denoted ${\cal C}_{p\_rtc}^{l}(\bar n)$
and ${\cal C}_{p\_rtc}^{u}(\bar n)$, and use all of these bounds to
compute bounds on the ratio. A lower (resp. upper) bound on the ratio
is given by $\frac{{\cal C}_{p\_rtc}^{l}(\bar n)}{{\cal
    C}_{p}^{u}(\bar n)}$ (resp. $\frac{{\cal C}_{p\_rtc}^{u}(\bar
  n)}{{\cal C}_{p}^{l}(\bar n)}$). Both bounds define an inferred
(safely approximated) band for the actual ratio, which is compared
with the (intended) ratio given in the specification (the band $[0,
  2*A]$) to produce the verification outcome as explained above.

\secbeg
\subsection{Using the accumulated cost for detecting hot spots}
\label{sec:accumulated-cost}
\secend

So far, we have used the standard notion of cost in the examples for
simplicity. However, in our approach we also use the \emph{accumulated
  cost}~\cite{gen-staticprofiling-iclp16-short}, inferred by \ciaopp,
to detect which of the run-time check predicates (properties) have a
higher impact
on the overall \overhead, and are thus promising targets for
optimization.
Given space restrictions we provide just the main idea and an example.
The \emph{accumulated cost} is based on the notion of
\emph{cost centers}, which in our approach are predicates to which
execution costs are assigned during the execution of a program. The
programmer can declare which predicates will be cost centers.
Consider again a predicate $p$, and its instrumented version $p\_rtc$
that performs run-time checks, and let ${\cal C}_p(\bar n)$ and ${\cal
  C}_{p\_rtc}(\bar n)$ be their corresponding standard cost
functions. Let $ck$ represent a run-time check predicate (e.g.,
\kbd{list/1}, \kbd{num/1}, \kbd{var/1}, etc.).  Let
$\Diamond_{p\_rtc}$ be the set of run-time check predicates used by
$p\_rtc$. Assume that we declare that the set of cost centers to be
used by the analysis, $\Diamond$, is $\Diamond_{p\_rtc} \cup
\{p\_rtc\}$. In this case, the cost of a (single) call to $p\_rtc$
\emph{accumulated in} cost center $ck$, denoted ${\mathcal
  C}_{p\_rtc}^{ck}(\bar n)$, expresses how much of the standard cost
${\cal C}_{p\_rtc}(\bar n)$ is attributed to run-time check $ck$
predicate (taking into account all the generated calls to $ck$).  The
$ck$ predicate with the highest ${\mathcal C}_{p\_rtc}^{ck}(\bar n)$
is a hot spot, and thus, its optimization can be more profitable to
reduce the overall \overhead.  The predicate $ck$ with the highest
${\mathcal C}_{p\_rtc}^{ck}(\bar n)$ is not necessarily the most
costly by itself, i.e., the one with the highest standard cost. For
example, a high ${\mathcal C}_{p\_rtc}^{ck}(\bar n)$ can be caused
because $ck$ is called very often.  We create a ranking of run-time
check predicates according to their accumulated cost. This can help in
deciding which assertions and properties to simplify/optimize first to
meet an overhead target.

Since $p\_rtc$ is declared as a cost center, the overall, absolute
\overhead
can be computed as $\sum_{ck \in \Diamond_{p\_rtc}} {\mathcal
  C}_{p\_rtc}^{ck}(\bar n)$. In addition ${\cal C}_{p\_rtc}(\bar n) =
\sum_{q \in \Diamond} {\mathcal C}_{p\_rtc}^{q}(\bar n)$, and ${\cal
  C}_p(\bar n) = {\mathcal C}_{p\_rtc}^{p\_rtc}(\bar n)$. Thus, we
only need to infer accumulated costs and combine them to both detect
hot spots and compute the \kbd{rtc\_ratio} described in
Section~\ref{sec:express-admissible-overhead}.

\begin{example}[Detecting hot spots]
\label{ex:detecting-hotspots}
\emph{%
Let \kbd{app\_rtc/3} denote the instrumented version for run-time
checking of predicate \kbd{app/3} in Example~\ref{ex:running-code}.
The following table shows the cost centers automatically declared by
the system, which are the predicate \kbd{app\_rtc/3} itself and the
run-time checking properties it uses (first column), as well as the
accumulated costs of a call to \kbd{app\_rtc(A,B,\_)} in each of
those cost centers, where $l_X$ represents the length of list $X$
(second column):}

\vspace*{1mm}
\begin{small}
\begin{tabular}{r|l}
\textbf{Cost center ($ck$)} & \textbf{${\mathcal C}_{app\_rtc}^{ck}(l_A, l_B)$}  \\
\cline{1-2}
\kbd{app\_rtc/3}    & $l_A + 1$ \\
\kbd{list/1} & $3 \times (l_A - 1)^2+6 \times (l_A+1) \times (l_B+1) + 8 \times (l_A + 1) -12$  \\
\kbd{var/1}   & $l_A + 1$ \\
\kbd{bit\_ops/1}     & $3 \times (l_A + 1)$ \\
\end{tabular}
\end{small}

\noindent\emph{
It is clear that the \emph{hot spot}
is the \kbd{list/1} property, which is the responsible of the
change in complexity order of the instrumented version
\kbd{app\_rtc/3} from linear 
to quadratic.}
\end{example}


\begin{table}[t]
\begin{tabular}{r|l}
\kbd{app(A,B,\_)}    & list concatenation \\
\kbd{oins(E,L,\_)}   & insertion into an ordered list\\
\kbd{mmtx(A,B,\_)}   & matrix multiplication \\
\kbd{nrev(L,\_)}     & list reversal \\
\kbd{ldiff(A,B,\_)}  & 2 lists difference \\
\kbd{sift(A,\_)}     & sieve of Eratosthenes \\
\kbd{pfxsum(A,\_)}   & sum of prefixes of a list of numbers \\
\kbd{bsts(N,T)}      & membership checks in a binary search tree \\
\end{tabular}
\caption{Description of the benchmarks.}
\label{fig:benchmarks}
\end{table}

\secbeg
\section{Implementation and Experimental Evaluation}
\label{sec:exp-results}
\secend

We have implemented a prototype of our approach within the \ciao system,
using \ciaopp's abstract interpretation-based resource usage analysis
and its combined static and dynamic verification framework.
Table~\ref{fig:benchmarks} contains a list of the benchmarks that we
have used in our experiments.\footnote{Sources available at
\url{http://cliplab.org/papers/rtchecks-cost/}.}
Each benchmark has assertions with properties related to shapes,
instantiation state, variable freeness, and variable sharing, as well
as in some cases more complex properties such as, for examples,
sortedness.  
The benchmarks and assertions were chosen to be simple enough to have
easily understandable costs but at the same time produce interesting
cost functions and overhead ratios.

As stated throughout the paper, our objective is to exploit static cost
analysis to obtain guarantees on program performance and detect cases
where adding run-time checks introduces overhead
that is not admissible.
%
To this end, we have considered the code instrumentation 
scenarios discussed previously, i.e.\ (cf.\
Examples~\ref{ex:running-rtchecks-orig} and
\ref{ex:running-rtchecks-orig-interface}): \\
\vspace{1mm}
\begin{tabular}{rcc}
\textsf{performance}
& \textsf{static assertion checking}
& \textsf{run-time checking instrumentation}        \\
\cline{1-3}
Original & no                   & no (\textsf{off}) \\
Worst    & no                   & yes (\textsf{full})\\
Base     & \emph{shfr + eterms} & yes (\textsf{opt})\\
\end{tabular}
\vspace{1mm}
\noindent
and we have performed for each benchmark and each scenario run-time
checking overhead analysis and verification, following the proposed
approach. The optimization in the \textsf{opt} case has been performed
using the \emph{eterms+shfr} domains.
The resource inferred in these experiments is the number of resolution
steps (i.e., each clause body is assumed to have unitary cost).
The experiments were performed on a MacBook Pro with 2.5GHz Intel Core
i5 CPU, 10 GB 1333 MHz DDR3 memory, running macOS Sierra
10.2.6.

Tables~\ref{tbl:cost-expressions-exact} and
\ref{tbl:cost-expressions-ub} show the results that our prototype
obtains for the different benchmarks.
In Table~\ref{tbl:cost-expressions-exact} we group the benchmarks for
which the analysis is able to infer the exact cost function, while in
Table~\ref{tbl:cost-expressions-ub} we have the benchmarks for which
the analysis infers a safe upper-bound of their actual resource
consumption.
The analysis also infers lower bounds, but we do not show them and
concentrate instead on the upper bounds for conciseness. Note that in
those cases where the analysis infers exact bounds
(Table~\ref{tbl:cost-expressions-exact}), the inferred lower and upper
bounds are of course the same.  Column~\benchmarks{} shows the name of
the entry predicate for each benchmark. 
Column~\colrtc{} indicates the scenario, as defined before, i.e., 
no run-time checks (\textsf{off}); full run-time checks
(\textsf{full}); or only those left after optimizing via static
verification (\textsf{opt}).


\begin{table}[t]
  \renewcommand{\times}{\cdot}
  \renewcommand{\cfrac}{\tfrac}
  \scriptsize
  \caption{Experimental results (benchmarks for which analysis infers exact cost functions).}
  \label{tbl:cost-expressions-exact}
  \setlength\tabcolsep{0.1pt}
  \renewcommand{\arraystretch}{2}
\begin{tabular}{|r|c|p{5cm}|r|r|c|c|}
\cline{1-7}

\benchmarks & \colrtc & \centering{\colexp} & \coldev  & \coltimes & \colover & \colaover  \\
  \hhline{|=|=|=|=|=|=|=|}

  \multirow{3}{*}{\texttt{app(A,B,\_)}}
       & \textsf{off} & $l_A+1$ & $0.0 $ 
       & \fmtFloatTwo{98.13} & &  \\ \cline{2-7}
       & \textsf{full} & ${l_A}^2 + 6 \times l_A \times l_B + 17 \times  l_A + 6 \times  l_B + 8 $ & $0.0 $ 
       & \fmtFloatTwo{521.18} & $l_A + l_B$ & \textsf{false} \\ \cline{2-7}
       & \textsf{opt} & $ 3 \times l_A + 2 \times l_B + 8 $  & $0.0 $
       & \fmtFloatTwo{311.98} & $\cfrac{l_B}{l_A} + 1$ & \textsf{false} \\
  \hhline{|=|=|=|=|=|=|=|}
  \multirow{3}{*}{\texttt{nrev(L,\_)}}
       & \textsf{off}& $ \tfrac{1}{2} \times {l_L}^2 + \tfrac{3}{2} \times l_L + 1$ & $0.0 $ 
       & \fmtFloatTwo{218.15} & &  \\ \cline{2-7}
       & \textsf{full} & $ \tfrac{1}{2} \times {l_L}^3 + 7 \times {l_L}^2 + \tfrac{29}{2} \times {l_L} + 8  $  & $0.0 $ 
       & \fmtFloatTwo{885.08} & $l_L$ & \textsf{false} \\ \cline{2-7}
       & \textsf{opt} & $   \tfrac{1}{2} \times {l_L}^2 + \tfrac{5}{2} \times  {l_L} + 7 $  & $0.0 $
       & \fmtFloatTwo{756.82} & $1$ & \textsf{checked} \\
    \hhline{|=|=|=|=|=|=|=|}
      \multirow{3}{*}{\texttt{sift(A,\_)}}
          & \textsf{off} & $ \tfrac{1}{2} \times {l_A}^2 + \tfrac{3}{2} \times {l_A} + 1 $ & $0.0 $ 
          & \fmtFloatTwo{255.55} & &  \\ \cline{2-7}
          & \textsf{full} & $ \tfrac{2}{3} \times {l_A}^3 + \tfrac{15}{2} \times {l_A}^2 + \tfrac{95}{6} \times {l_A} + 7$ & $0.0 $ 
          & \fmtFloatTwo{980.63} & $ l_A $ & \textsf{false} \\ \cline{2-7}
          & \textsf{opt} & $\tfrac{1}{2} \times {l_A}^2  + \tfrac{7}{2} \times {l_A} + 7$  & $0.0 $
          & \fmtFloatTwo{521.65}  & $1$ & \textsf{checked} \\
  \hhline{|=|=|=|=|=|=|=|}
      \multirow{3}{*}{\texttt{pfxsum(A,\_)}}
          & \textsf{off} & $ l_A + 2 $ & $0.0 $ 
          & \fmtFloatTwo{146.98}  & &  \\ \cline{2-7}
          & \textsf{full} & $ 2 \times {l_A}^2 + 12 \times l_A + 14 $ & $0.0 $ 
          & \fmtFloatTwo{749.94} & $ l_A $ & \textsf{false} \\ \cline{2-7}
          & \textsf{opt} & $ 3 \times {l_A} + 10 $  & $0.0 $
          & \fmtFloatTwo{469.71} & $1$ & \textsf{checked} \\
  \hhline{|=|=|=|=|=|=|=|}
\end{tabular}
\vspace{-7mm}
\end{table}


Column~\colexp{} shows the resource usage functions inferred by our
resource analysis, for each of the cases. These functions depend on
the input data sizes of the entry predicate (as before, $l_X$
represents the length of list $X$).
In order to measure the precision of the functions inferred, in
Column~\coldev{} we show the average deviation of the bounds obtained
by evaluating the functions in Column~\colexp{}, with respect to the
costs measured with dynamic profiling. The input data for dynamic
profiling was selected to exhibit worst case executions. In those
cases where the inferred bounds is exact, the deviation is always
$0.0\%$. In Column~\colover{} we show the relative \overhead as the ratio
(\kbd{rtc\_ratio})
between the complexity order of the cost of the instrumented code
(for \textsf{full} or \textsf{opt}), and the complexity
order of the cost corresponding to the original code (\textsf{off}).
Finally, in Column~\coltimes{} we list the \emph{cost} analysis time
for each of the three cases.\footnote{%
  This time does not include the static analysis and verification time
  in the \textsf{opt} case, performed with the \emph{eterms+shfr}
  domains, since the process of simplifying at compile-time the
  assertions is orthogonal to this paper. Recent experiments and
  results on this topic can be found in~\cite{optchk-journal-scp}.}

\begin{table}[ht]
  \renewcommand{\times}{\cdot}
  \renewcommand{\cfrac}{\tfrac}
  \scriptsize
  \caption{Experimental results (rest of the benchmarks; we show the upper bounds).}
  \label{tbl:cost-expressions-ub}
  \setlength\tabcolsep{0.1pt}
  \renewcommand{\arraystretch}{2}
\begin{tabular}{|r|c|p{7.3cm}|r|r|c|c|}
\cline{1-7}

\benchmarks & \colrtc & \centering{\colexp} & \coldev  & \coltimes & \colover & \colaover  \\
  \hhline{|=|=|=|=|=|=|=|}

  \multirow{3}{*}{\texttt{oins(E,L,\_)}}
       & \textsf{off} & $ l_L + 2 $ & $0.09 $  
       & \fmtFloatTwo{142.55}  & &   \\ \cline{2-7}
       & \textsf{full} & $ \tfrac{1}{3} \times {l_L}^3 + \tfrac{9}{2} \times {l_L}^2 - \tfrac{5}{2} \times l_L + \tfrac{11}{3} $ & $99.93 $ 
       & \fmtFloatTwo{917.39}  & $ {l_L}^2  $ & \textsf{false}  \\ \cline{2-7}
       & \textsf{opt*} & $\tfrac{3}{2} \times {l_L} + 6$  & $50.14 $
       & \fmtFloatTwo{340.15} & $ 1 $ & \textsf{checked} \\
  \hhline{|=|=|=|=|=|=|=|}

  \multirow{3}{*}{\texttt{mmtx(A,B,\_)}}
       & \textsf{off} & $ {r_A} \times {c_A} \times {c_B} + 3 \times {r_A} \times {c_B} + 2 \times {r_A} - 2 \times {c_B} $ & $7.58 $   
       & \fmtFloatTwo{460.21} & &  \\ \cline{2-7}
                & \textsf{full} &  $ 4 \times {r_A}^2 {c_A} \times {c_B} + 4 \times {r_A}^2 \times {c_A} + 4 \times {r_A}^2 \times {c_B} + 4 \times {r_A}^2 + {r_A} \times {c_A}^2 \times {c_B} + 4 \times {r_A}  \times {c_A}^2 + 2 \times {r_A} \times {c_A} \times {c_B}^2 + 11 \times {r_A} \times {c_A} \times {c_B} + 20 \times {r_A} \times {c_A} + 15 \times {r_A} + 7 $ & $0.0 $ 
                & \fmtFloatTwo{1682.54} & $ N^\dagger $ & \textsf{false} \\ \cline{2-7}
       & \textsf{opt} & $ {r_A} \times {c_A} \times {c_B} + 2 \times {c_A} \times {c_B} + 2 \times {r_A} \times {c_A} + 4 \times {r_A} \times {c_A} + 6  \times {r_A}  + 2 \times {c_A} + 11$
       & $0.0 $
       & \fmtFloatTwo{1120.23}  & $ 1 $ & \textsf{checked} \\
  \hhline{|=|=|=|=|=|=|=|}

  \multirow{3}{*}{\texttt{ldiff(A,B,\_)}}
          & \textsf{off} & $ l_A \times l_B + 2 \times l_A + 1 $ & $2.06$ 
          & \fmtFloatTwo{786.22}  & &  \\ \cline{2-7}
          & \textsf{full}  & $   {l_A}^2 + 3 \times {l_A} \times {l_B} + 10 \times {l_A} + 2 \times {l_B} + 7  $ & $0.27 $ 
          & \fmtFloatTwo{1769.22} & $\tfrac{l_A}{l_B} + 1$ & \textsf{false}  \\ \cline{2-7}
          & \textsf{opt} & ${l_A} \times {l_B} + 5 \times {l_A} + 2 \times {l_B} + 8$  & $0.0 $
          & \fmtFloatTwo{1226.15} &  $1$ & \textsf{checked} \\
  \hhline{|=|=|=|=|=|=|=|}
    \multirow{3}{*}{\texttt{bsts(N,T)}}
       & \textsf{off} & $ d_T + 3 $ & $0.1 $ 
       & \fmtFloatTwo{714.83}  & &  \\ \cline{2-7}
       & \textsf{full} & $ 3 \times  2^{(d_T + 2)} + 3 \times  2^{(d_T + 1)} + 3 \times  2^{(d_T - 1)} + 3 \times  2^{d_T} + \tfrac{3}{2} \times  (d_T - 1)^2 + \tfrac{47}{2} \times  (d_T + 2) - \tfrac{27}{2} $ & $1.19 $ 
       & \fmtFloatTwo{438.72} & $ \dfrac{{2}^{d_T}}{d_T} $ & \textsf{false}  \\ \cline{2-7}
       & \textsf{opt*} & $3 \times 2^{({d_T} + 1)} +   4 \times {d_T} +  14 $  & $4.01 $
       & \fmtFloatTwo{245.09} & $ \dfrac{{2}^{d_T}}{d_T} $ & \textsf{false} \\
  \hhline{|=|=|=|=|=|=|=|}
\end{tabular}

\begin{minipage}{\textwidth}  
\scriptsize 
$\dagger N = max(r_A,c_A,c_B)$ 
\end{minipage}
\vspace{-9mm}
\end{table}

From the results shown in Column~\colover we see that the analysis
correctly detects that the 
full run-time checking versions of the benchmarks (\textsf{full} case)
are asymptotically worse than the original program, showing for
example a quadratic asymptotic ratio (run-time checking overhead)
for \texttt{oins/3}, or even
exponential for \texttt{bsts/2}. In the case of \texttt{app/3}, we can
see that the asymptotic relative overhead is linear, but the 
instrumented versions become dependent on the size of both arguments, while
originally the cost was only depending on the size of the first list (though
probably it is worth it since
the list check
on the second argument should be
performed anyway). On the
other hand, for all the benchmarks, except for \texttt{app/3} and
\texttt{bsts/2}, the resulting asymptotic relative overhead of the optimized
run-time checking version (\textsf{opt} case), is null, i.e.,
$\overheadacronym = 1$.

In the case of \texttt{bsts/2}, the overhead is still exponential because
the type analysis is not able to statically prove the property
\emph{binary search tree}. Thus, it is still necessary to traverse the
input binary tree at run-time in order to verify it. However, the
optimized version traverses the input tree only once, while the full
version traverses it on each call, which is reflected in the resulting
cost function. In any case, note that the exponential functions are on
the depth of the tree $d_T$, not on the number of nodes.
Analogously, in \texttt{oins/3} the static analysis is
not able to prove the \emph{sorted} property for the input list,
although in that case the complexity order does not change for the
optimized version, only increasing the constant coefficients of the
cost functions.
We have included optimized versions of these two cases (marking them
with \texttt{*}) to show the change in the overhead if the properties
involved were verified; however, the \emph{eterms+shfr} domains used
cannot prove these complex properties.

Column~\colaover{} shows the result of verification 
(i.e., \texttt{checked}/\texttt{false}/\texttt{check}) 
assuming a global
assertion for all predicates in all the benchmarks stating that the
relative \overhead 
should not be larger than 1 ($\overheadacronym \leq
1$). Finally, Column~\coltimes{} shows that the analysis time is
$\approx 4$ times slower on versions with full instrumentation, and
$\approx 2$ times slower on versions instrumented with run-time checks
after static analysis, respectively, but in any case all analysis
times are small.

We believe that these results are encouraging and strongly suggest
that our approach can provide information that can help the programmer
understand statically, at the algorithmic level whether the overheads
introduced by the run-time checking required by the assertions in the
program are acceptable or not.


\secbeg
\section{Conclusions}
\label{sec-discuss-and-concl}
\secend

We have proposed a method that uses static analysis to infer bounds on
the overhead that run-time checking introduces in programs.  The
bounds are functions parameterized by input data sizes.  Unlike
profiling, this approach can provide guarantees for all possible
execution traces, and allows assessing how the overhead grows as the
size of the input grows. We have also extended the Ciao assertion
verification framework to express ``admissible'' overheads, and
statically and automatically check whether the instrumented program
conforms with such specifications.  Our experimental evaluation
suggests that our method is feasible and also promising in providing
bounds that help the programmer understand at the algorithmic
level the overheads introduced by the run-time checking required for
the assertions in the program, in different scenarios, such as
performing full run-time checking or checking only the module
interfaces.


\begin{small}
\secbeg
\bibliographystyle{acmtrans}
\bibliography{../../../../clip/bibtex/clip/clip,../../../../clip/bibtex/clip/general}

\end{small}

\end{document}